\documentclass[10pt,prc,aps,superscriptaddress]{revtex4}
\usepackage{graphicx}
\tighten
\begin{document}
\draft
\title{Towards order-by-order calculations of the nuclear and neutron matter \\
equations of state in chiral effective field theory}

\author{F. Sammarruca}
\affiliation{Department of Physics, University of Idaho, Moscow, ID 83844, USA}

\author{L. Coraggio}
\affiliation{Istituto Nazionale di Fisica Nucleare, \\
Complesso Universitario di Monte S. Angelo, Via Cintia - I-80126 Napoli, Italy}

\author{J. W. Holt}
\affiliation{Department of Physics, University of Washington,
Seattle, WA 98195, USA}

\author{N. Itaco}
\affiliation{Istituto Nazionale di Fisica Nucleare, \\
Complesso Universitario di Monte  S. Angelo, Via Cintia - I-80126 Napoli, Italy}
\affiliation{Dipartimento di Fisica, Universit\`a di Napoli Federico II, \\
Complesso Universitario di Monte  S. Angelo, Via Cintia - I-80126 Napoli, Italy}

\author{R. Machleidt}
\affiliation{Department of Physics, University of Idaho, Moscow, ID 83844, USA}

\author{L. E. Marcucci}
\affiliation{Dipartimento di Fisica ``Enrico Fermi'', Universit\`a
di Pisa, Largo Bruno Pontecorvo 3 - I-56127 Pisa, Italy}
\affiliation{Istituto Nazionale di Fisica Nucleare, Sezione di Pisa,\\
Largo Bruno Pontecorvo 3 - I-56127 Pisa, Italy}

\date{\today}
\begin{abstract}
We calculate the nuclear and neutron matter equations of state from microscopic nuclear 
forces at different orders in chiral effective field theory and with varying momentum-space cutoff 
scales. We focus attention on how the order-by-order convergence depends on the choice of 
resolution scale and the implications for theoretical uncertainty estimates on the isospin 
asymmetry energy. Specifically we study the equations of state using consistent NLO and 
N$^2$LO (next-to-next-to-leading order) chiral potentials where the low-energy constants $c_D$ 
and $c_E$ associated with contact vertices in the N$^2$LO chiral three-nucleon force are fitted 
to reproduce the binding energies of $^3$H and $^3$He as well as the beta-decay lifetime of 
$^3$H. At these low orders in the chiral expansion there is little sign of convergence, while 
an exploratory study employing the N$^3$LO two-nucleon force together with the N$^2$LO 
three-nucleon force give first indications for (slow) convergence with low-cutoff potentials and
poor convergence with higher-cutoff potentials. The consistent NLO and N$^2$LO potentials 
described in the present work provide the basis for estimating theoretical uncertainties 
associated with the order-by-order convergence of nuclear many-body calculations in chiral 
effective field theory.
\end{abstract}
\pacs {21.65.+f, 21.30.Fe} 
\maketitle 
        
\section{Introduction} 
\label{Intro} 
The equation of state (EoS) of highly neutron-rich matter is important
for understanding wide ranging questions in contemporary nuclear
structure physics, from the structure of rare isotopes (e.g., the
thickness of neutron skins) to the properties of neutron stars. 
A quantity of central importance in many of these phenomena
is the density-dependent nuclear symmetry energy, arising as the
difference between the energy per particle of symmetric nuclear matter
and pure neutron matter at a given density.

In astrophysical contexts, the EoS of neutron-rich matter is required
over many orders of magnitude in the nuclear density, potentially up
to ten times that of saturated nuclear matter. 
In principle, both one-boson-exchange as well as
Dirac-Brueckner-Hartree-Fock (DBHF) calculations can access the
high-density regime of the EoS. 
However, boson-exchange models (see, for instance,
Ref.~\cite{Catania}) typically employ three-nucleon forces (3NFs) with
little connection to the associated nucleon-nucleon ($NN$) force. For
instance, large 3NF contributions arise from $\Delta$-isobar
intermediate states, whereas explicit $\Delta$ isobars are missing
from the $NN$ part. 
Relativistic approaches to the nuclear matter problem have been
centered around the DBHF scheme \cite{FS14}. 
The main strength of this framework is in its ability to account for
an important class of 3NFs, namely virtual nucleon-antinucleon
excitations, that lead to nuclear matter saturation close to the
empirical density and energy \cite{sammarruca12}.

Nowadays, microscopic nuclear many-body theory typically starts from the
low-energy realization  of QCD, chiral effective field theory
\cite{Wei68,Wein79}, and fits unresolved  nuclear dynamics at short
distances to the properties of two- and  few-nucleon systems alone. 
The resulting potentials are then used to make predictions in nuclear
many-body systems. The chiral effective field theory approach to nuclear 
and neutron matter has succeeded in producing realistic equations of 
state only with the inclusion of repulsive 3NFs arising at order N$^2$LO 
(next-to-next-to-leading order) 
in the chiral power counting \cite{coraggio13,krueger,coraggio14,wellenhofer14}. 
In chiral effective field theory, the dominant two-pion-exchange
component of this 3NF, with associated $c_{1,3,4}$ low-energy
constants, is constructed consistently with the N$^2$LO two-body
force. 
Many-body perturbation theory with low-momentum chiral nuclear forces
\cite{coraggio14} has been shown to reproduce qualitatively the
saturation behavior found with renormalization-group-evolved two-body
forces with refit low-energy constants in the 3NF sector
\cite{bogner05,hebeler11}. However, all low-momentum interactions are 
limited in calculations of the EoS to densities where the characteristic 
momentum scale (on the order of the Fermi momentum) is below the scale 
set by the momentum-space cutoff $\Lambda$
in the $NN$ potential regulating function, which for chiral $NN$ forces
typically has the form:
\begin{equation}
f(p',p) = \exp[-(p'/\Lambda)^{2n} - (p/\Lambda)^{2n}] \; ,
\label{reg}
\end{equation}
where $\Lambda \lesssim 500$\,MeV is associated with the onset of
favorable perturbative properties. 
Nonperturbative methods for computing the EoS with chiral nuclear 
forces are under active investigation
\cite{gezerlis13,hagen14,roggero14,wlazlowski14,carbone14} and may 
allow for higher values of the cutoff $\Lambda$ (but still remaining below the 
chiral breakdown scale of about 1 GeV). Furthermore, concerning the range 
of densities which can be reliably accessed, we note that the Fermi 
momentum must be lower than the cutoff scale, regardless of the nature 
(perturbative or nonperturbative) of the many-body calculations.
Although designed to reproduce similar $NN$ scattering phase shifts, $NN$ 
potentials with different regulator functions will yield different predictions in the 
nuclear many-body problem due to their different off-shell behavior.  On the 
other hand, appropriate re-adjustment of the low-energy constants that appear 
in the nuclear many-body forces is expected to reduce the dependence on the 
regulator function \cite{coraggio13}.

Estimates of theoretical uncertainties \cite{furnstahl15} for calculations of the 
equation of state have largely focused on varying the low-energy constants 
and resolution scale at which nuclear dynamics are probed 
\cite{bogner05,hebeler11,gezerlis13,krueger,coraggio13,coraggio14}. In the 
present work we lay the foundation for order-by-order calculations of nuclear 
many-body systems by presenting consistent NLO and N$^2$LO chiral nuclear 
forces whose relevant short-range three-body forces are fit to $A=3$ binding 
energies and the lifetime of the triton. We then assess the accuracy with which 
infinite nuclear matter properties and the isospin asymmetry energy can be 
predicted from order-by-order calculations in chiral effective field theory. 
Identifying the dominant sources of uncertainty in nuclear many-body calculations 
is an important open problem, especially as more stringent constraints on the EoS
of neutron-rich matter and its density dependence are becoming available 
\cite{Tsang+}. In computing the EoS, we employ the nonperturbative 
particle-particle ladder approximation, which re-sums an important class of 
diagrams accounting for Pauli-blocking in the medium. 

%
We will employ $NN$ potentials at NLO, N$^2$LO and N$^3$LO in
the chiral expansion at resolution scales in the range $450\,{\rm MeV}
\le \Lambda \le 600$\,MeV (for a recent review, see Ref.~\cite{ME11}). Note 
that we omit discussion of leading order (LO) chiral potentials, which
are very crude and substantially less quantitative than interactions from 
sixty years ago, such as the Gammel-Thaler potential \cite{Gammel}. Thus, predictions at 
LO are not expected to meaningfully add to the discussion. 
Beyond NLO we include the leading chiral 3NF, whose low-energy
constants are fitted to reproduce the binding energies of $^3$H and
$^3$He as well as the beta-decay lifetime of $^3$H~\cite{Marc}. 
Definite conclusions on convergence will be limited to the third 
order (N$^2$LO) in the chiral power counting, where fully 
consistent two- and three-nucleon forces are currently available. 
Similar studies, limited to pure neutron matter
at and below the nuclear saturation density, have been performed in
Ref.~\cite{gezerlis13} up to N$^2$LO with two-body forces alone and in
Ref.~\cite{krueger} for N$^2$LO and N$^3$LO chiral nuclear forces.

The paper is organized as follows. In Section \ref{Descr}, we will
describe detailed features of the NLO, N$^2$LO and N$^3$LO chiral $NN$
potentials employed in the present work, together with consistent
N$^2$LO three-nucleon forces when appropriate. 
In Section \ref{BHF} we outline the calculations of the energy per
particle of symmetric nuclear matter and pure neutron matter in the
particle-particle ladder approximation employing chiral nuclear forces
at different chiral orders and resolution scales.
Results for the EoS and the nuclear symmetry energy up to a density of
$\rho \simeq 0.3$\,fm$^{-3}$ are presented. 
We end with a summary and conclusions.

\section{Description of the calculations} 
\label{Descr} 
\subsection{The two-nucleon potentials} 
\label{NN} 

\begin{table}
\centering
\begin{tabular}{|c||c|c|c|c|c|}
\hline
NLO & $\Lambda$ (MeV) & $n$& $c_1$ & $c_3$ & $c_4$ \\
\hline     
 &  450 & 2&       &       &       \\
 &  500 & 2&       &       &       \\
 &  600 & 2&       &       &       \\
\hline
\hline
N$^2$LO & $\Lambda$ (MeV) & $n$& $c_1$ & $c_3$ & $c_4$ \\
\hline     
 & 450 & 3& -0.81 & -3.40 & 3.40  \\
 & 500 & 3& -0.81 & -3.40 & 3.40  \\
 & 600 & 3& -0.81 & -3.40 & 3.40  \\
\hline
\hline
N$^3$LO & $\Lambda$ (MeV) & $n$& $c_1$ & $c_3$ & $c_4$ \\
\hline     
 & 450 & 3& -0.81 & -3.40 & 3.40  \\
 & 500 & 2& -0.81 & -3.20 & 5.40  \\
 & 600 & 2& -0.81 & -3.20 & 5.40  \\
\hline
\hline
\end{tabular}
\caption{Values of $n$ and low-energy constants of the dimension-two $\pi N$ Lagrangian, 
$c_{1,3,4}$, at each order and for each type of cutoff in the regulator function given in 
Eq.~(\ref{reg}). None of the $c_i$'s appears at NLO. The low-energy constants are given in 
units of GeV$^{-1}$.}
\label{tab1}
\end{table}

In the present investigation we consider $NN$ potentials at order $(q/\Lambda_\chi)^2$,
$(q/\Lambda_\chi)^3$ and $(q/\Lambda_\chi)^4$ in the chiral power counting, where 
$q$ denotes the small scale set by external nucleon momenta or the pion mass and 
$\Lambda_\chi$ is the chiral symmetry breaking scale. Chiral $NN$ potentials at NLO
and N$^2$LO, corresponding to $(q/\Lambda_\chi)^2$ and $(q/\Lambda_\chi)^3$, 
have been constructed previously in Ref.~\cite{NLO} for cutoffs ranging from $\Lambda$
= 450 MeV to about 800 MeV. With varying chiral order and cutoff scale, the low-energy 
constants in the two-nucleon sector are refitted to elastic $NN$ scattering 
phase shifts and properties of the deuteron. The low-energy constants $c_{1,3,4}$ 
associated with the $\pi \pi N N$ contact couplings of the ${\cal L}^{(2)}_{\pi N}$ chiral
Lagrangian are given in Table~\ref{tab1}.
We note that the $c_{i}$ can be extracted from $\pi N$ or $NN$ scattering data. The potentials we use here
\cite{EM03,ME11} follow the second path. At N$^2$LO, taking the range determined 
in $\pi N$ analyses as a starting point, values were chosen to best reproduce $NN$ data at that order, 
see Table 2 of Ref.~\cite{ME11}. 
At N$^3$LO, high-precision required a stronger adjustment of $c_4$ depending on the regulator 
function and cutoff. The fitting procedure is discussed in Ref.~\cite{ME11}, where it is noted that 
the larger value for $c_4$ has, overall, a very small impact but lowers the $^3F_2$ phase shift 
for a better agreement with the phase shift analysis.

\begin{figure}[!t] 
\vspace*{-1cm}
\hspace*{-1.5cm}
\scalebox{0.50}{\includegraphics{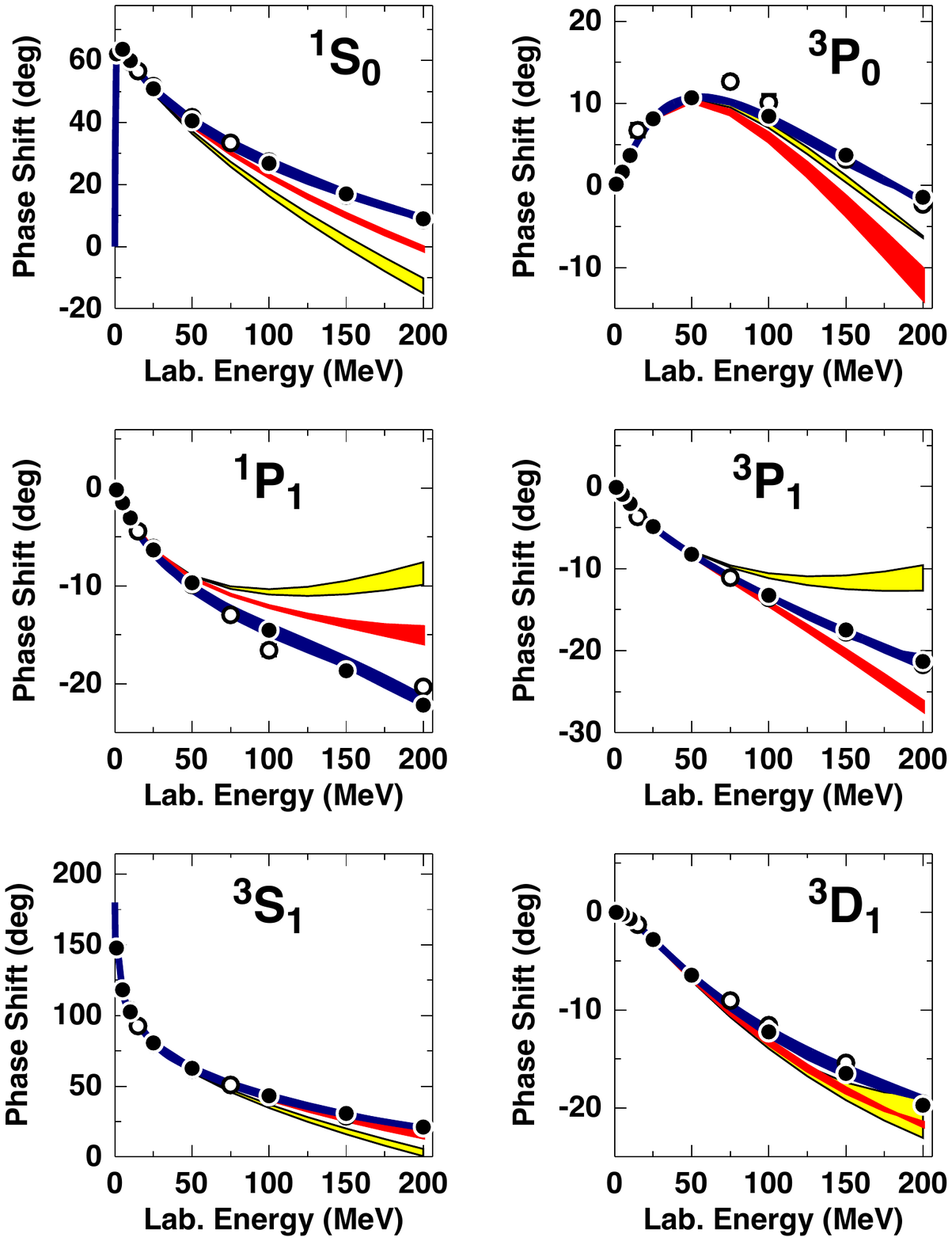}}
\hspace*{-2.5cm}
\scalebox{0.50}{\includegraphics{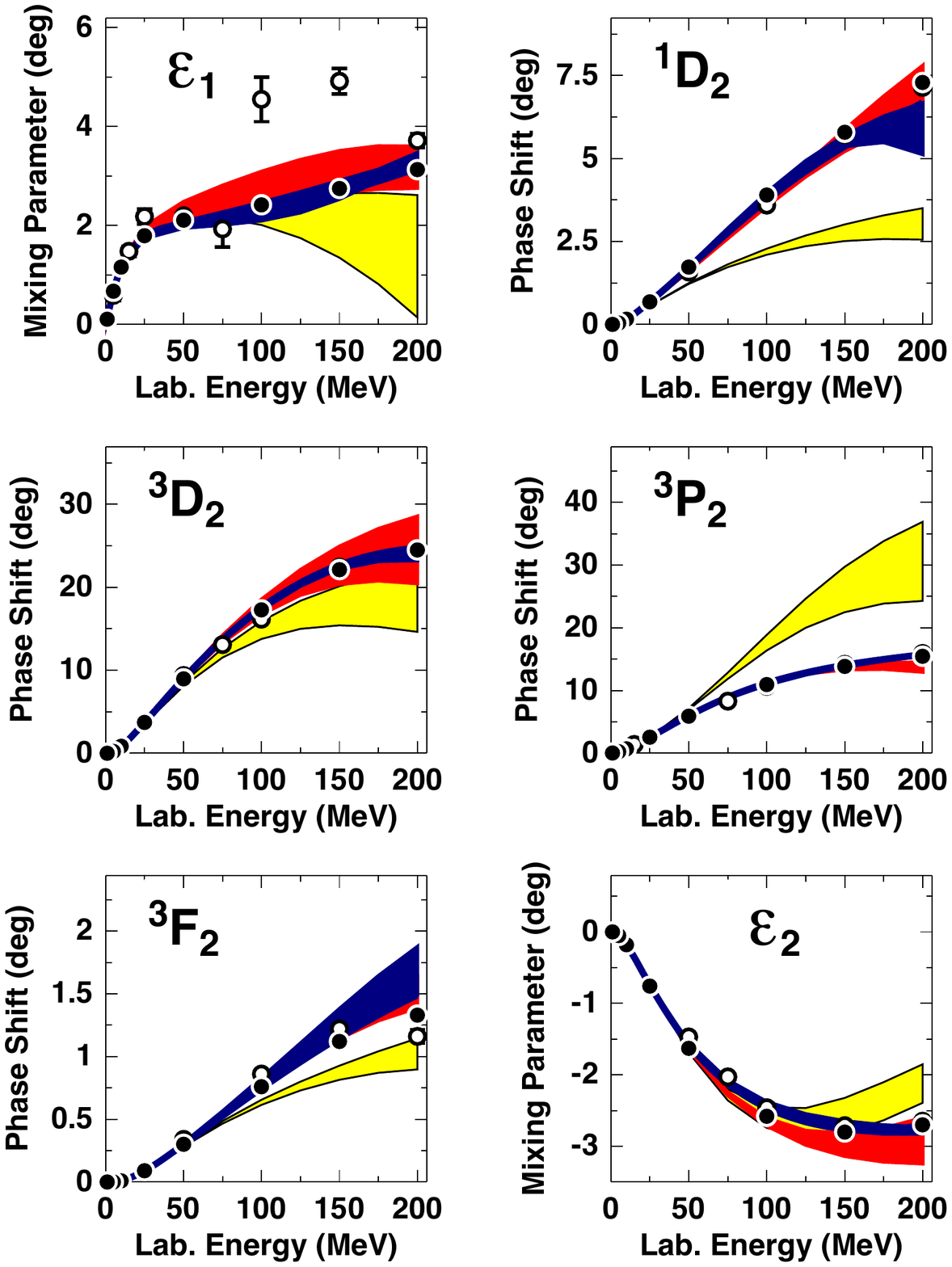}}
\vspace*{-2.5cm}
\caption{(Color online) Phase shifts for some selected $NN$ partial waves. The 
yellow, red, and blue bands show the variations of the predictions with changing cutoffs 
between 450 and 600 at  NLO, N$^2$LO, and N$^3$LO, respectively.}
\label{ph}
\end{figure}

In Ref.~\cite{NLO}, it was found that the two-body scattering phase shifts can be 
described well at NLO up to a laboratory energy of about 100 MeV, while the N$^2$LO 
potential fits the data up to 200 MeV. Interestingly, in the latter case the $\chi^2/$datum was 
found to be essentially cutoff independent for variations of $\Lambda$ between 450 
and approximately 800 MeV. Finally, we also use $NN$ potentials constructed 
at next-to-next-to-next-to-leading order (N$^3$LO) \cite{ME11,EM03}, with low-energy 
constants $c_{1,3,4}$ as displayed in Table~\ref{tab1}. However, at N$^3$LO, $NN$ 
potentials with cutoffs up to 800 MeV are not available.  Therefore, in the present study, we 
limit the cutoff range to 450-600 MeV.

Before proceeding to the nuclear and neutron matter calculations, we demonstrate the 
dependence of $NN$ scattering phase shifts on the chiral order and on the
choice of the cutoff scale in the regulating function Eq.~(\ref{reg}). Results are shown
in Fig.~\ref{ph}, where the yellow, red and blue bands indicate the 
NLO, N$^2$LO, and N$^3$LO results, respectively, obtained from varying the 
cutoff between 450 and 600 MeV. Although N$^2$LO calculations can achieve sufficient
accuracy in selected partial wave channels up to $E_{\rm lab} = 200$\,MeV, only the
N$^3$LO interactions achieve the level of high-precision potentials, characterized by a 
$\chi^2/$datum $\sim 1$.

At the two-body level each time the chiral order is increased, the $NN$ contact terms 
and/or the two-pion-exchange contributions proportional to the low-energy constants
$c_{1,3,4}$ are refitted. We recall that at N$^2$LO no new $NN$ contact terms
are generated, and therefore improved cutoff independence in the $NN$ phase 
shifts (compare the yellow and red bands in Fig.~\ref{ph}) is due to changes in the 
two-pion-exchange contributions. At N$^2$LO, subleading $\pi \pi N N$ vertices 
enter into the chiral $NN$ potential. These terms encode the important physics of 
correlated two-pion-exchange and the excitation of intermediate $\Delta$(1232) 
isobar states. Thus, only at this order is it possible to obtain a realistic description of the 
$NN$ interaction at intermediate-range, traditionally generated through the exchange of 
a fictitious $\sigma$ meson of intermediate mass. At N$^3$LO in the chiral power 
counting, the 15 additional $NN$ contact terms (bringing the total to 24 at N$^3$LO) 
result in a much improved description of $NN$ scattering phase shifts.

We observe that the calculated phase shifts are in most cases not renormalization group 
invariant, though with increasing chiral order the dependence on the cutoff scale
is generally reduced. The standard Weinberg's power counting in which contributions
to the $NN$ potential are computed perturbatively with loop integrals renormalized through 
counterterms, is employed in the present work. Weinberg's scheme implicitly assumes that 
the counterterms introduced to renormalize the perturbative potential are sufficient to also 
renormalize its nonperturbative resummation, e.g., in the Lippmann-Schwinger equation for 
computing phase shifts. Kaplan {\it et al.} \cite{Kaplan}, however, pointed out the presence 
of problems with this assumption, which stimulated intense discussion in the 
literature~\cite{PC}. In particular, Nogga {\it et al.}~\cite{NTK} performed a systematic 
investigation of Weinberg's power counting at lowest order and proposed a modified scheme                
in which contact terms from NLO are promoted to LO to take care of the cutoff dependence 
in $^3P_0$ and $^3P_2$, and from N$^3$LO to LO to address the same problem in 
$^3D_2$. On the other hand, the consistency problem of Weinberg's power counting
appears to be minimal when implemented together with finite-cutoff regularization below 
the high-energy scale of the effective field theory. It has been 
shown \cite{NLO} that, at a given order, cutoff 
ranges can be identified where the $\chi ^2$ of the fit to the $NN$ data is essentially flat
(that is, cutoff independent). In other words, order-by-order renormalization can be 
accomplished successfully with finite cutoffs in Weinberg's power counting, and
errors associated with variations in the momentum-space cutoff (a regularization 
prescription that does not respect chiral symmetry) are typically of the same order as the 
error expected from the truncated chiral expansion.

\subsection{Chiral three-body interactions} 
\label{3bf} 

\begin{figure}[!t] 
\centering
\includegraphics{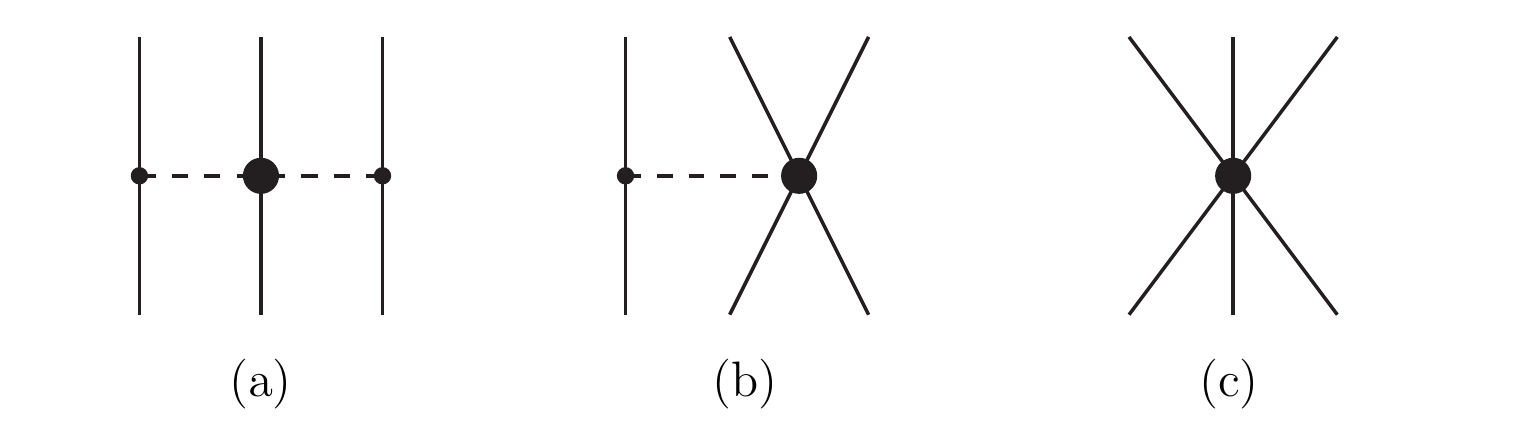}
\caption{Diagrams of the 3NF at N$^2$LO. See text for more details.} 
\label{3b}
\end{figure}

Three-nucleon forces make their appearance at third order in the chiral power 
counting. They are expressed as the sum of three contributions: the long-range 
two-pion-exchange part with $\pi\pi NN$ vertex proportional to the low-energy
constants $c_1,c_3,c_4$, the medium-range one-pion exchange diagram
proportional to the low-energy constant $c_D$, and finally the short-range contact 
term proportional to $c_E$. The corresponding diagrams are shown in 
Fig.~\ref{3b}, labeled as (a), (b), (c), respectively.

The natural framework to include 3NFs in the particle-particle ladder
approximation to the energy per particle in homogeneous nuclear matter would be the Bethe-Faddeev equation. 
To facilitate this inclusion, we employ the
density-dependent $NN$ interaction derived in Refs.~\cite{holt09,holt10} from
the N$^2$LO chiral three-body force. This effective interaction is obtained by
summing one particle line over the occupied states in the Fermi sea. Neglecting
small contributions \cite{hebeler10} from terms depending on the center-of-mass 
momentum, the resulting $NN$ interaction can be expressed in analytical
form with operator structures identical to those of free-space $NN$ interactions.
For symmetric nuclear matter all three-body forces contribute, while for pure
neutron matter only terms proportional to the low-energy constants $c_1$ and
$c_3$ are nonvanishing \cite{holt10,hebeler10}. Previous studies (see e.g., 
Ref.\ \cite{coraggio14}) have found that 3NF contributions to 
the energy per particle are dominant at the Hartree-Fock level. When 
three-body forces are approximated with a density-dependent $NN$ interaction,
certain topologies are missing at and beyond second order in perturbation theory.
In the case of a pure contact interaction the additional topologies reduce the 
second-order contribution to the energy per particle by about 50\% at saturation
density \cite{kaiser12}, which corresponds in the present calculation to an 
uncertainty of $\Delta E / A \lesssim 1$\,MeV at saturation density.

We fix the low-energy constants $c_D$ and $c_E$ that appear in the N$^2$LO 3NF 
within the three-nucleon sector. Specifically, we constrain them to reproduce 
binding energies of $A=3$ nuclei together with the Gamow-Teller matrix element in tritium 
$\beta$-decay, following a well established procedure \cite{Marc,Gar06,Gaz09,Viv13,Pia13,Mar13,Viv14}.
The values of $c_D$ and $c_E$ are given in Table~\ref{tab2} for the different chiral 
orders and cutoff scales. We note that the values at N$^3$LO in 
Table~\ref{tab2} are extracted from Refs.\ \cite{Marc,coraggio14}, while those at N$^2$LO
have been computed in the present work. Although efforts are in progress to incorporate 
potentially important N$^3$LO 3NF contributions 
\cite{ishikawa07,Ber08,Ber11}, both in the fitting procedure and in the neutron and nuclear 
matter equations of state presented in the following section, the current ``N$^3$LO'' study 
is limited to the inclusion of the N$^2$LO three-body force together with the N$^3$LO 
two-body force, an approximation that is commonly used in the literature but whose 
associated uncertainties have not been carefully analyzed. In Refs.~\cite{krueger,Tews}, 
calculations of the neutron matter energy per particle at N$^3$LO show a small effect 
(of about -0.5 MeV) at saturation density for the potentials of our purview~\cite{ME11}.
The Hartree-Fock contributions to the energy per particle of symmetric nuclear matter 
from the N$^3$LO 3NF are attractive and on the order of 7 MeV at 
saturation density \cite{krueger}. The inclusion of 3NFs at N$^3$LO, however, necessitates 
a refitting of the $c_D$ and $c_E$ low-energy constants, which has not yet been performed 
and would likely result in a smaller change to the total energy per particle at saturation
density. Most recently, evidence has been reported~\cite{Heb2015} that sub-leading terms 
in the 3NF may provide important contributions to the triton binding energy, as well as 
indications that similar conclusions apply in symmetric nuclear matter.

\begin{figure}[!t] 
\centering
\includegraphics[width=8cm,height=6cm]{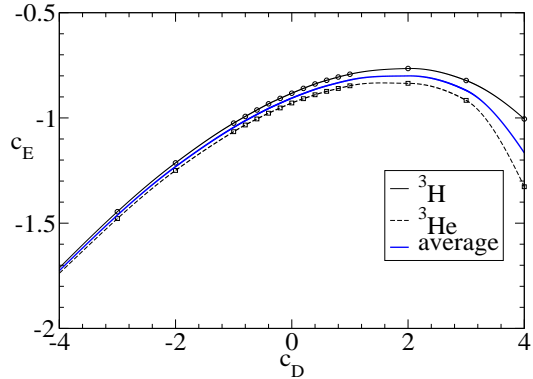}
\caption{(Color online) $c_E$-$c_D$ trajectories fitted to reproduce the experimental
$^3$H and $^3$He binding energies in the case of the 
N$^2$LO $NN$ interaction plus 3NF with $\Lambda=600$ MeV. The average
curve (in blue) is also displayed.} 
\label{fig:n2lo600}
\end{figure}

In the case of the N$^2$LO $NN$ interaction with $\Lambda=600$ MeV,
small charge-symmetry-breaking effects have emerged in the
fitting procedure. This is visible in Fig.~\ref{fig:n2lo600}, where the
$c_D$-$c_E$ trajectories which reproduce the experimental
$^3$H and $^3$He binding energies are displayed. Allowing for 
charge-symmetry breaking, the values for $c_E$ are -0.833 and -0.885 
for $^3$H and $^3$He, respectively. The value shown in Table~\ref{tab2}
is the average of these two. The error in the $A=3$ binding energies,
when the average value of $c_E$ is used, is $\sim 40$ keV. We do not know
at present the origin of this (small) charge-symmetry breaking effect.

\begin{table}                
\centering
\begin{tabular}{|c||c|c|c|}
\hline
N$^2$LO & $\Lambda$ (MeV) & $c_D$ & $c_E$ \\
\hline     
 & 450 &-0.326 & -0.149 \\
 & 500 &-0.165 & -0.169   \\
 & 600 & 0.456 & -0.859   \\
\hline
\hline
N$^3$LO & $\Lambda$ (MeV) & $c_D$ & $c_E$ \\
\hline
 & 450 &-0.24  &-0.11    \\
 & 500 & 0.0   & -0.18   \\
 & 600 & -0.19 & -0.83  \\
\hline
\end{tabular}
\caption                                                    
{Values of the $c_D$ and $c_E$ low-energy constants 
obtained using the N$^2$LO 3NF in conjunction with 
$NN$ interactions of different orders, for several values of the 
cutoff $\Lambda$. These constants do not appear at 
NLO.} 
\label{tab2}
\end{table}

\section{Nuclear and neutron matter calculations: results and discussion} 
\label{BHF} 


In this section we present results for the symmetric nuclear matter and neutron 
matter equations of state employing the particle-particle ladder approximation with 
the $NN$ and $3N$ forces described above. In the traditional hole-line expansion \cite{day67}, the
particle-particle ladder diagrams comprise the leading-order
contributions.
The next set of diagrams is comprised of the three hole-line contributions, which includes the 
third-order particle-hole (ph) diagram considered in Ref.~\cite{coraggio14}. The 
third-order hole-hole (hh) diagram (fourth order in the hole-line expansion) was found to 
give a negligible contribution to 
the energy per particle at normal density irrespective of the cutoff (see Table 
II and Table III of Ref.~\cite{coraggio14}). The ph diagram is relatively much larger, 
bringing in an uncertainty of about $\pm$1 MeV (accounting for cutoff dependence
$\Lambda \simeq 400 - 500$\,MeV) on 
the potential energy per particle at normal density. 

It is insightful to compare these values 
with those from Refs.~\cite{cc1,cc2}. In Ref.~\cite{cc1}, the authors report on coupled-cluster 
calculations in symmetric nuclear matter including pp and hh diagrams (as well as an exact 
treatment of the Pauli operator). The overall effect, as seen from comparing the 
first and last entries in Table II of Ref.~\cite{cc1}, is very small around saturation density, 
consistent with Table II in Ref.~\cite{coraggio14}, and grows to 1.5 MeV at the highest 
Fermi momentum included in the study. Note that these calculations adopt the N$^3$LO 
potential (with $\Lambda$=500 MeV) and only two-nucleon forces. On the other hand, in 
Ref.~\cite{cc2} coupled-cluster calculations in nucleonic matter were performed at N$^2$LO 
with two- and three-body forces and with the inclusion of selected triples clusters, namely 
correlations beyond pp and hh ladders. The effect of these contributions is found to be 
negligible in neutron matter and about 1 MeV per nucleon in symmetric matter in the density
range under consideration~\cite{cc2}. Furthermore, it was shown in earlier studies \cite{Baldo}
with meson-theoretic interactions (with larger momentum-space cutoffs) that when 
additional three hole-line contributions are taken into account, large cancellations occur which 
result in a small net effect on the energy per particle. This is especially the case when the 
continuous choice is adopted for the auxiliary potential~\cite{Baldo}. In summary, we 
conclude that a realistic estimate of the impact of using a nonperturbative approach beyond 
pp correlations is about $\pm$1 MeV in nuclear matter around saturation density and much 
smaller in neutron matter. As we show below, such uncertainties are significantly smaller than 
those associated with variations in the cutoff scale.

\begin{figure}[!t]
\vspace*{-3cm}
\includegraphics[width=8.7cm]{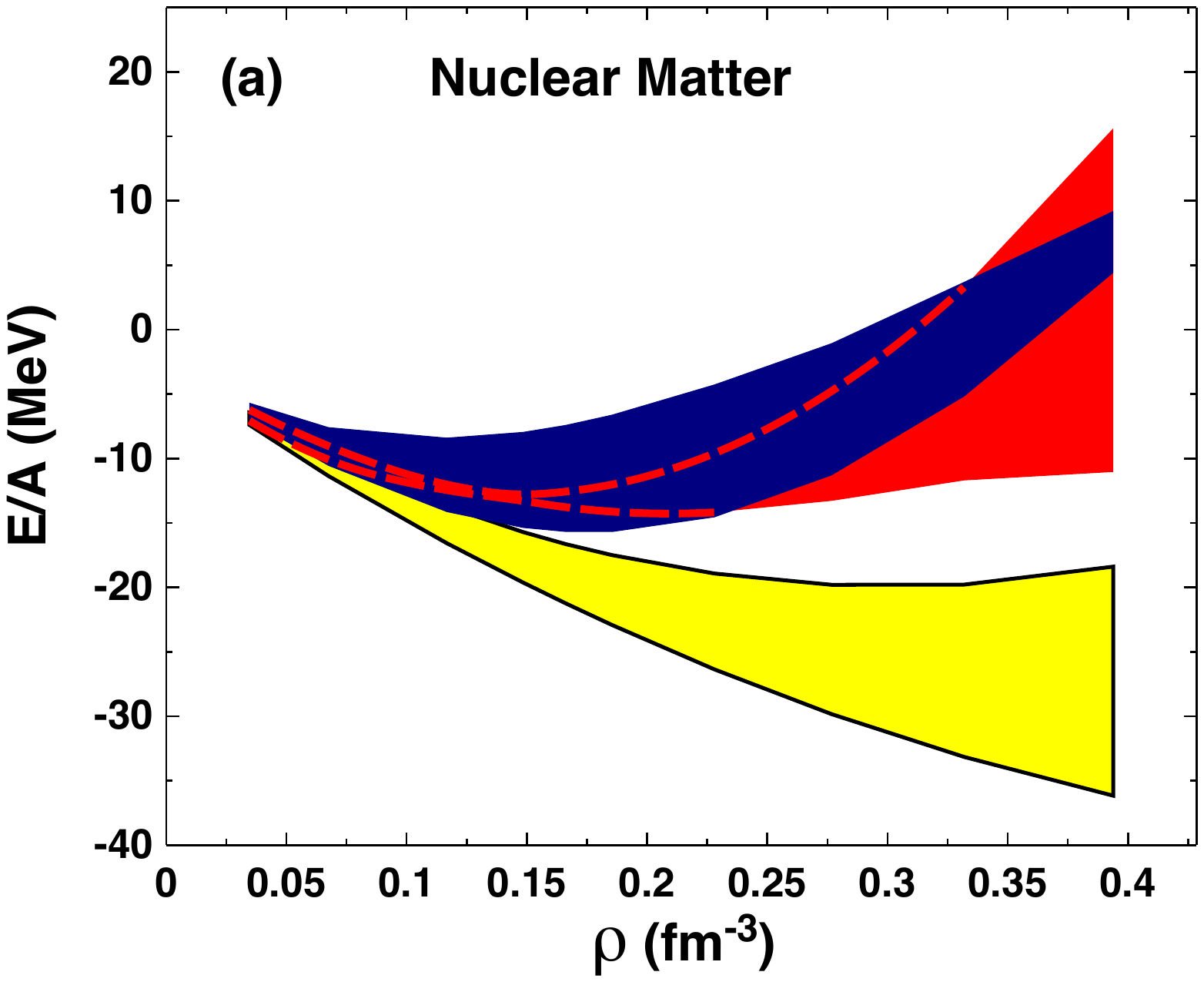}\hspace{.1in}
\includegraphics[width=8.7cm]{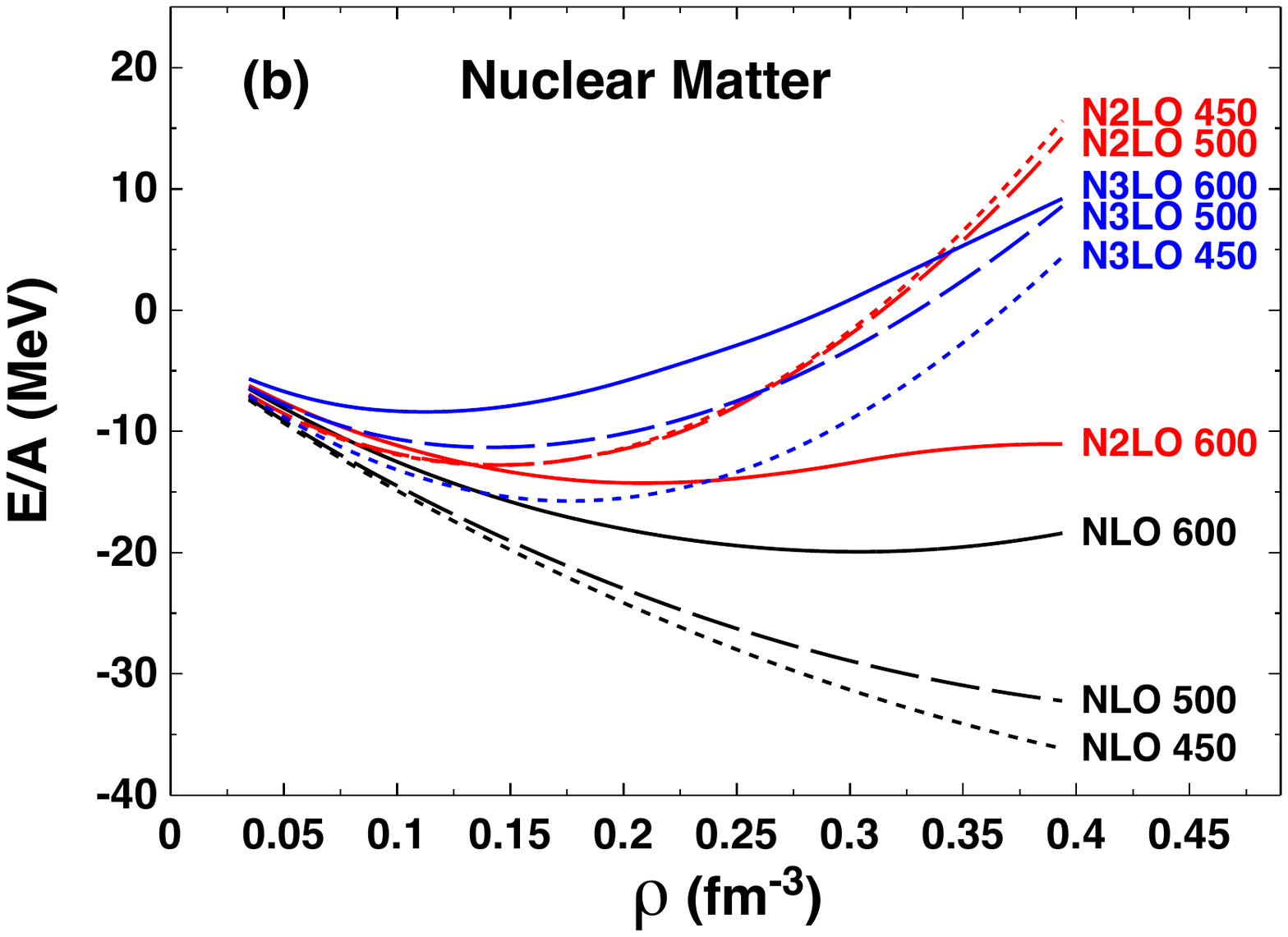}\hspace{.1in}
\vspace*{-3cm}
\caption{(Color online) Energy/nucleon (E/A) in symmetric nuclear matter
as a function of density, $\rho$. Left frame: The yellow and
red bands represent the uncertainties in the predictions due to cutoff variations
as obtained in complete calculations at NLO and N$^2$LO, respectively.
The blue band is the result of a calculation employing N$^3$LO $NN$ potentials together with
N$^2$LO 3NFs. 
The dashed lines show the upper or lower limits of hidden bands. Right frame: predictions at 
the specified order and cutoff value.} 
\label{snm}
\end{figure}

Our results for the energy per particle as a function of the nuclear density are shown in 
Fig.~\ref{snm} for symmetric nuclear matter. We note that the particle-particle ladder 
approximation employed in the present work is in good agreement with the perturbative 
results available at N$^3$LO from Ref.~\cite{coraggio14} including up to third-order pp 
diagrams. In the left panel of Fig.~\ref{snm}, 
the shaded bands in yellow and red represent the spread of our complete calculations 
conducted at NLO and N$^2$LO, respectively. The blue band is the result of a calculation 
that employs N$^3$LO $NN$ potentials together with N$^2$LO 3NFs. In all cases shown, 
the cutoff is varied over the range 450-600 MeV. As noted before, the N$^3$LO 3NFs and 
4NFs are at present omitted, and the resulting convergence pattern gives an estimate on
the theoretical uncertainty of the calculation (and not of the chiral effective field theory
expansion per se). On the right-hand side of the figure, the individual curves 
corresponding to each order and cutoff are displayed. We observe that at NLO the 
potentials constructed at lower cutoff scales do not exhibit saturation until very high 
densities. On the other hand, for the 600 MeV cutoff potential the $^1S_0$ partial wave 
(together with the $^3S_1$ partial wave) is sufficiently repulsive to enable saturation at a 
relatively smaller density. We observe that the convergence pattern for the low-cutoff 
($\Lambda = 450 - 500$\,MeV) potentials is significantly better than for the 600 MeV 
potential. Overall there is a large spread from cutoff variations both at NLO and N$^2$LO beyond 
nuclear matter saturation density. Moreover, the bands at these two orders do not overlap, 
suggesting that their width is not a suitable representation of the uncertainty. Although the
(incomplete) N$^3$LO calculation reveals a strong reduction of the cutoff dependence, it 
is important to notice that an uncertainty of about 8 MeV remains at saturation density.
While we do not expect much of a change in nuclear matter predictions from 4NFs
\cite{4BF1,4BF2,krueger}, it is quite possible that the inclusion of N$^3$LO 3NFs might reduce 
either the cutoff dependence or improve the convergence pattern. This will be an interesting 
subject for future investigations.

\begin{figure}[!t]
\vspace*{-3cm}
\includegraphics[width=8.7cm]{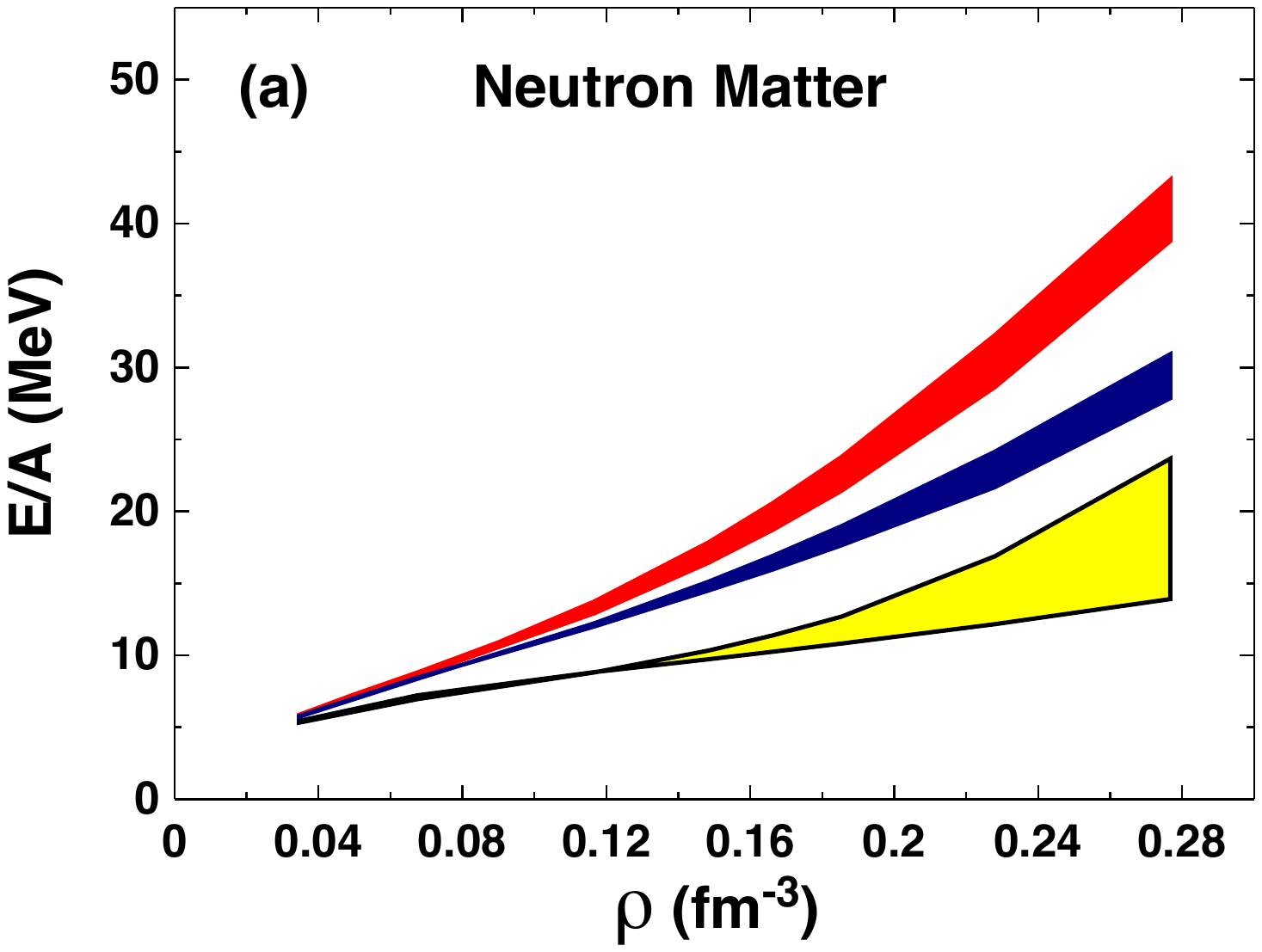}\hspace{.1in}
\includegraphics[width=8.7cm]{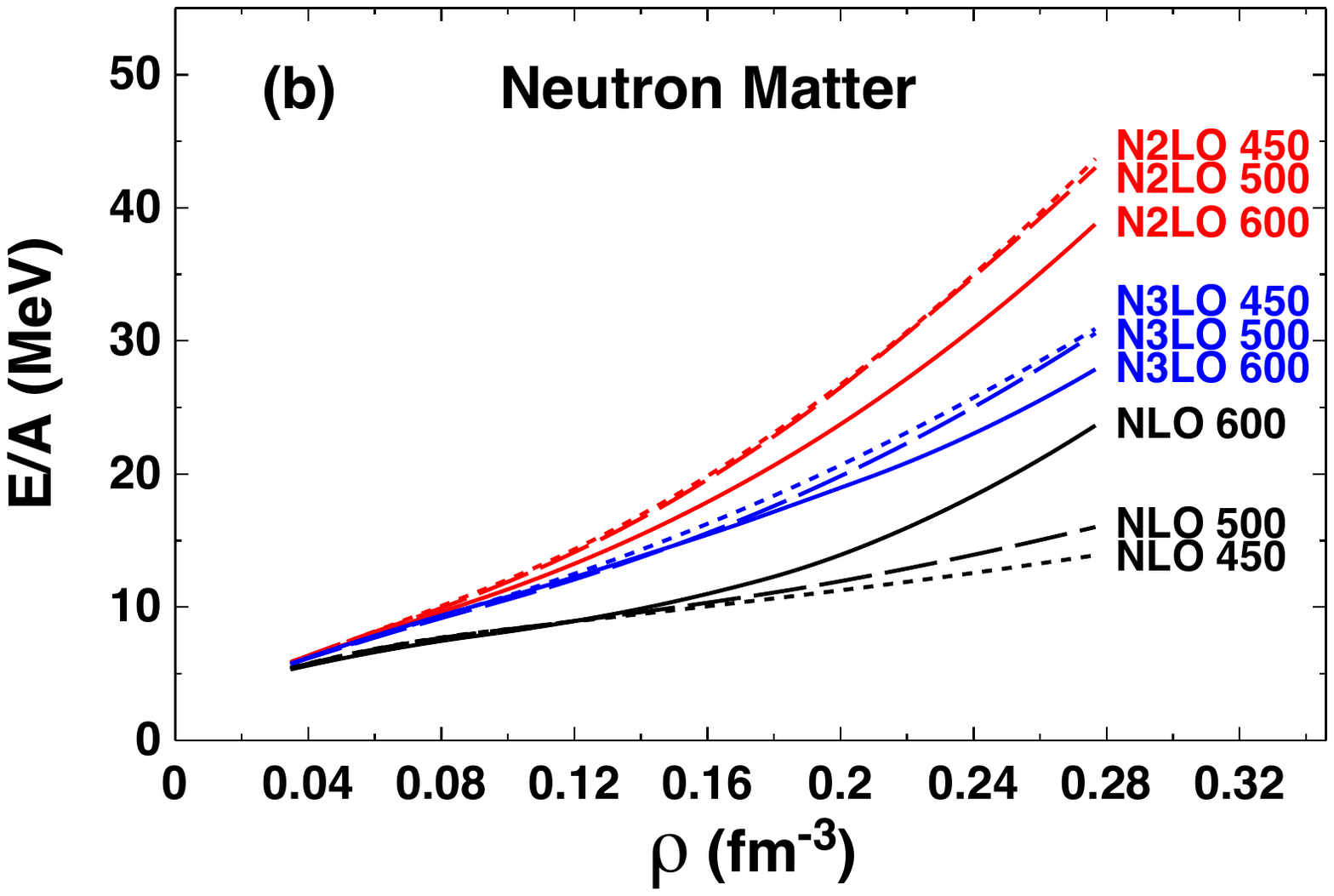}\hspace{.1in}
\vspace*{-3cm}
\caption{(Color online) As in Fig.~\ref{snm} for pure neutron matter. 
} 
\label{nm}
\end{figure}

The results for neutron matter are presented in Fig.~\ref{nm}, where the left and right 
panels have the same meaning as in Fig.~\ref{snm}. Note that the range of densities 
under consideration is smaller for neutron matter in order to keep the Fermi momentum 
below the cutoff in all cases.  We see a large spread at NLO for the largest densities considered, 
whereas the band has only moderate size at the next order and remains small for our 
N$^3$LO calculation. Similar to what was observed in symmetric nuclear matter, the bands at NLO and N$^2$LO 
do not overlap in neutron matter. In addition the N$^3$LO band does not generally overlap
with the N$^2$LO band. Therefore, the variation obtained by changing the cutoff does not 
seem to provide a reliable representation of the uncertainty at the given order. A better way to 
estimate such uncertainty is to consider the difference between the predictions at two consecutive 
orders.

\begin{figure}[!t] 
\centering         
\vspace*{-3.5cm}
\scalebox{0.45}{\includegraphics{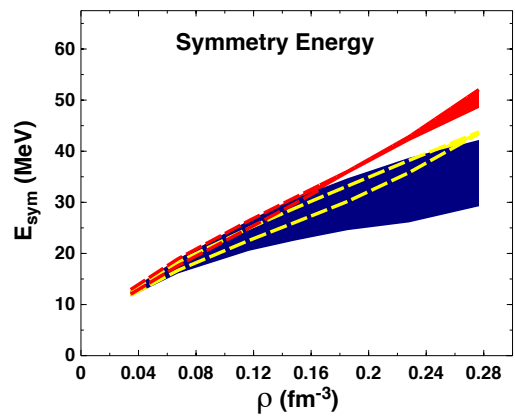}}
\vspace*{-3.5cm}
\caption{(Color online) The symmetry energy, $E_{\rm sym}$, as a function of density, $\rho$.
Meaning of bands and dashed lines as in the right panel of Fig.~\ref{snm}.} 
\label{esym}
\end{figure}

In Fig.~\ref{esym} we present the results for the symmetry energy $E_{\rm sym}$, which
is defined as the strength of the quadratic term in an expansion of the energy per particle 
in asymmetric matter with respect to the asymmetry parameter $\alpha$:                                   
\begin{equation}
\bar E (\rho,\alpha) \approx \bar E (\rho,\alpha = 0) + E_{\rm sym}\alpha^2 + {\cal{O}}(\alpha ^4) \; , 
\end{equation}
where $\bar E = E/A$ is the energy per particle and $\alpha= (\rho_n - \rho_p)/(\rho_n+\rho_p)$. 
The nearly linear behavior of $\bar E(\rho,\alpha)$ with $\alpha ^2$ has been confirmed by many 
microscopic calculations (see for instance Ref.~\cite{bombaci91} and more recently
Refs.~\cite{AS,Drischler}). It justifies the common approximation of neglecting powers beyond
$\alpha ^2$ in the expansion above and thus defining 
the symmetry energy as the difference between the energy per particle in neutron
matter and symmetric nuclear matter.

It is well known that $E_{\rm sym}$ enters crucially in discussions of nuclear stability, and 
its density dependence around normal density strongly correlates with the neutron skin 
thickness of nuclei and the radius of (low-mass) neutron stars. As mentioned in Sec.\ \ref{Intro}, 
systematic efforts are ongoing to set better empirical constraints on the symmetry energy, 
through both laboratory and astrophysical measurements. It is therefore important to have 
an understanding of the theoretical uncertainty affecting calculations of this quantity.
The spread due to the change of the cutoff values in our NLO, N$^2$LO, 
and N$^3$LO calculations is represented by the three bands as before. As         
observed previously for symmetric matter, the spread due to cutoff variations            
remains large at N$^2$LO, with some minimal overlap with the NLO band.         
The N$^3$LO band reflects the large cutoff sensitivity previously 
observed in symmetric matter. Again, we conclude that the spread generated 
by changing the cutoff does not in general provide a reliable estimate
of the theoretical uncertainty. 

We close this section with some information on the density dependence of the symmetry
energy, as revealed by the $L$ parameter,
\begin{equation}
L = 3 \rho_0 \Big ( \frac{\partial E_{\rm sym}}{\partial \rho} \Big )_{\rho_0} \;.
\label{L} 
\end{equation}
Namely, the $L$ parameter reflects the slope of the symmetry energy at saturation density $\rho_0$.
Our N$^3$LO result can be summarized as $L = 39.5 ^{+17.2} _{-13.6}$ MeV, whereas at N$^2$LO 
we find $L = 76.9 ^{+16.1} _{-31.2}$ MeV. We do not report a corresponding value at NLO, since, at
that order, only the $\Lambda$=600 MeV case show some (late) saturating behavior. 
Constraints on $L$ are not yet stringent, and can be quoted as $L= 70 \pm 25$ MeV \cite{Tsang+}.

\section{Conclusions}
\label{Concl} 

We have reported predictions for the energy per particle in symmetric nuclear matter and pure 
neutron matter, focusing on uncertainties related to order-by-order convergence. Compared
to the consistent NLO and N$^2$LO results for the equations of state, which themselves exhibit 
relatively little overlap even for a large spread of momentum-space cutoffs $\Lambda = 450 - 
600$\,MeV, the results from employing N$^3$LO two-nucleon and N$^2$LO 
three-nucleon forces imply non-negligible
uncertainties associated with missing higher-order terms in the chiral expansion.
We find that the uncertainty associated with the cutoff variation is generally larger in symmetric
nuclear matter than in pure neutron matter, but in the latter case we find that the results from
one chiral order to the next have little overlap. This suggests that further systematic studies of 
the order-by-order convergence should be performed, together with variations in the resolution 
scale and low-energy constants, to accurately estimate the complete theoretical uncertainties 
in chiral effective field theory predictions of nuclear many-body systems.
The two- and three-body potentials considered in the present work can serve as a basis for such 
future uncertainty estimates.

\section*{Acknowledgments}
Support from the U.S. Department of Energy Office of Science, 
Office of Basic Energy Science, under Award Nos.        
DE-FG02-03ER41270 and DE-FG02-97ER41014 is acknowledged. Part of the
results presented here have been obtained at the INFN-Pisa computer
center.



\begin{thebibliography} {100}

\bibitem{Catania} Z.H. Li, U. Lombardo, H.-J. Schulze, and W. Zuo, Phys. Rev. C {\bf 77}, 
034316 (2008).

\bibitem{FS14} F. Sammarruca, Int. J. Mod. Phys. E, {\bf 22}, 1330031 (2014), and 
references therein.

\bibitem{sammarruca12} F. Sammarruca,  B. Chen, L. Coraggio, N. Itaco,
  and R. Machleidt, Phys. Rev. C {\bf 86}, 054317 (2012).

\bibitem{Wei68} S. Weinberg, Phys. Rev. {\bf 166}, 1568 (1968).

\bibitem{Wein79} S. Weinberg, Physica {\bf 96A}, 327 (1979).

\bibitem{coraggio13} L. Coraggio, J. W. Holt, N. Itaco, R. Machleidt and F. Sammarruca, 
 Phys. Rev. C {\bf 87}, 014322 (2013).
 
\bibitem{krueger} T. Kr{\"u}ger, I. Tews , K. Hebeler, and A. Schwenk, Phys. Rev. C {\bf 88},
 025802 (2013).

\bibitem{coraggio14} L. Coraggio, J. W. Holt, N. Itaco, R. Machleidt, L. E. Marcucci, and 
 F. Sammarruca, Phys. Rev. C {\bf 89}, 044321 (2014).

\bibitem{wellenhofer14} C. Wellenhofer, J. W. Holt, N. Kaiser and W. Weise, Phys. Rev. C
 {\bf 89}, 064009 (2014).

\bibitem{bogner05} S. K. Bogner, A. Schwenk, R. J. Furnstahl, and A. Nogga,  Nucl. Phys. A 
 {\bf 763}, 59 (2005).

\bibitem{hebeler11} K. Hebeler, S. K. Bogner, R. J. Furnstahl, A. Nogga, and A. Schwenk, Phys. Rev. C {\bf 83}, 031301 (2011).

\bibitem{gezerlis13} A. Gezerlis, I. Tews, E. Epelbaum, S. Gandolfi, K. Hebeler, A. Nogga, and 
 A. Schwenk, Phys. Rev. Lett. {\bf 111}, 032501 (2013).

\bibitem{hagen14} G. Hagen, T. Papenbrock, A. Ekstr{\"o}m, K. A. Wendt, G. Baardsen, 
 S. Gandolfi, M. Hjorth-Jensen and C. J. Horowitz, Phys. Rev. C {\bf 89}, 014319 (2014).

\bibitem{roggero14} A. Roggero, A. Mukherjee, and F. Pederiva, Phys. Rev. Lett. {\bf 112}, 221103 (2014).

\bibitem{wlazlowski14} G. Wlaz{\l}owski, J. W. Holt, S. Moroz,
  A. Bulgac, and K. Roche, Phys. Rev. Lett. {\bf 113}, 182503 (2014).

\bibitem{carbone14} A. Carbone, A. Rios, and A. Polls, arXiv:1408.0717.

\bibitem{furnstahl15} R. J. Furnstahl, D. R. Phillips, and S. Wesolowski, J. Phys. G {\bf 42}, 034028 (2015).

\bibitem{Tsang+} M.B. Tsang {\it et al.}, Phys. Rev. C {\bf 86}, 015803 (2012).

\bibitem{Gammel} J.L. Gammel, R.M. Thaler, Phys. Rev. {\bf 107}, 291 (1957).

\bibitem{ME11} R. Machleidt and D.R. Entem, Phys. Rep. {\bf 503}, 1 (2011).

\bibitem{Marc} L.E. Marcucci, A. Kievsky, S. Rosati, R. Schiavilla, and M. Viviani, 
Phys. Rev. Lett. {\bf 108}, 052502 (2012).

\bibitem{NLO} E. Marji, A. Canul, Q. MacPherson, R. Winzer, Ch. Zeoli, D.R. Entem, and R. Machleidt, Phys. Rev. C {\bf 88}, 054002 (2013).

\bibitem{EM03} D.R. Entem and R. Machleidt, Phys. Rev. C {\bf 68}, 041001 (2003).

\bibitem{holt09} J. W. Holt, N. Kaiser, and W. Weise, Phys. Rev. C {\bf 79}, 054331 (2009).

\bibitem{holt10} J. W. Holt, N. Kaiser, and W. Weise, Phys. Rev. C {\bf 81}, 024002 (2010).

\bibitem{hebeler10} K. Hebeler and A. Schwenk, Phys. Rev. C {\bf 82}, 014314 (2010).

\bibitem{kaiser12} N. Kaiser, Eur. Phys. J. A {\bf 48}, 58 (2012).

\bibitem{Gar06} A. Gardestig and D.R. Phillips, 
Phys. Rev. Lett. {\bf 96}, 232301 (2006).

\bibitem{Gaz09} D. Gazit, S. Quaglioni, and P. Navratil,
Phys. Rev. Lett {\bf 103}, 102502 (2009).

\bibitem{Viv13} M. Viviani, L. Girlanda, A. Kievsky, and L.E. Marcucci, 
Phys. Rev. Lett. {\bf 111}, 172302 (2013).

\bibitem{Pia13} M.\ Piarulli, L. Girlanda, L.E. Marcucci, S. Pastore, R. Schiavilla, and M. Viviani, Phys. Rev. C {\bf 87}, 014006  (2013).

\bibitem{Mar13} L.E. Marcucci, R. Schiavilla, and M. Viviani, 
Phys. Rev. Lett. {\bf 110}, 192503 (2013).

\bibitem{Viv14} M.\ Viviani, A.\ Baroni, L.\ Girlanda, A.\ Kievsky, L.E.\ Marcucci, and R.\ Schiavilla,
Phys.\ Rev.\ C {\bf 89}, 064004 (2014).

\bibitem{ishikawa07} S. Ishikawa and M. R. Robilotta, Phys. Rev. C {\bf 76}, 014006 (2007).

\bibitem{Ber08} V. Bernard, E. Epelbaum, H. Krebs, and U.-G. Meissner, 
Phys. Rev. C {\bf 77}, 064004 (2008).

\bibitem{Ber11} V. Bernard, E. Epelbaum, H. Krebs, and U.-G. Meissner, 
Phys. Rev. C {\bf 84}, 054001 (2011).

\bibitem{Heb2015} K. Hebeler {\it et al.}, arXiv:1502.02977.                     

\bibitem{Tews} I. Tews, T. Kr{\"u}ger, K. Hebeler, and A. Schwenk, Phys. Rev. Lett. {\bf 110}, 032504 (2013).

\bibitem{4BF1} E. Epelbaum, Eur. Phys. J. A {\bf 34}, 197 (2007).

\bibitem{4BF2} D. Rozpedzik {\it et al.}, Acta Phys. Polon. B {\bf 37}, 2889 (2006).

\bibitem{Kaplan} D.B. Kaplan, M.J. Savage, and M.B. Wise, Nucl. Phys. {\bf B478}, 629 (1996); 
Phys. Lett. B {\bf 424}, 390 (1998); Nucl. Phys. {\bf B534}, 329 (1998). 

\bibitem{PC} See Refs.~[146-164] in Ref.~\cite{ME11}.                      

\bibitem{NTK} A. Nogga, R.G.E. Timmermans, and U. van Kolck, Phys. Rev. C {\bf 72}, 054006 (2005). 

\bibitem{day67} B. D. Day, Rev. Mod. Phys. {\bf 39}, 719 (1967).

\bibitem{cc1} G. Baardsen {\it et al.}, Phys. Rev. C {\bf 88}, 054312 (2013).

\bibitem{cc2} G. Hagen {\it et al.}, Phys. Rev. C {\bf 89}, 014319 (2014).

\bibitem{Baldo} H.Q. Song, M. Baldo, G. Giansiracusa, and U. Lombardo, Phys. Rev. Lett. {\bf 81} 1584 (1998).

\bibitem{bombaci91} I. Bombaci and U. Lombardo, Phys. Rev. C {\bf 44}, 1892 (1991).

\bibitem{AS} D. Alonso and F. Sammarruca, Phys. Rev. C {\bf 67}, 054301 (2003).

\bibitem{Drischler} C. Drischler {\it et al.}, Phys. Rev. C {\bf 89}, 025806 (2014). 

\end{thebibliography}
\end{document}